\documentclass[aps,preprint]{revtex4}%
\usepackage{amsfonts}
\usepackage{amsmath}
\usepackage{amssymb}
\usepackage{graphicx}%
\begin{document}
\preprint{ }
\title[Renormalized EE of holographic CFTs]{Topological terms, AdS$_{2n}$ gravity and renormalized Entanglement Entropy of
holographic CFTs}
\author{Giorgos Anastasiou}
\affiliation{Departamento de Ciencias F\'isicas, Universidad Andr\'es Bello, Sazi\'e 2212,
Piso 7, Santiago, Chile}
\author{Ignacio J. Araya}
\affiliation{Departamento de Ciencias F\'isicas, Universidad Andr\'es Bello, Sazi\'e 2212,
Piso 7, Santiago, Chile}
\author{Rodrigo Olea}
\affiliation{Departamento de Ciencias F\'isicas, Universidad Andr\'es Bello, Sazi\'e 2212,
Piso 7, Santiago, Chile}
\pacs{PACS number}

\begin{abstract}
We extend our topological renormalization scheme for Entanglement Entropy to
holographic CFTs of arbitrary odd dimensions in the context of the AdS/CFT
correspondence. The procedure consists in adding the Chern form as a boundary
term to the area functional of the Ryu-Takayanagi minimal surface. The
renormalized Entanglement Entropy thus obtained can be rewritten in terms of
the Euler characteristic and the AdS curvature of the minimal surface. This
prescription considers the use of the Replica Trick to express the
renormalized Entanglement Entropy in terms of the renormalized gravitational
action evaluated on the conically-singular replica manifold extended to the
bulk. This renormalized action is obtained in turn by adding the Chern form as
the counterterm at the boundary of the $2n$-dimensional asymptotically AdS
bulk manifold. We explicitly show that, up to next-to-leading order in the
holographic radial coordinate, the addition of this boundary term cancels the
divergent part of the Entanglement Entropy. We discuss possible applications
of the method for studying CFT parameters like central charges.

\end{abstract}
\volumeyear{ }
\startpage{1}
\endpage{2}
\maketitle
\tableofcontents

\section{Introduction}

In Ref.\cite{AAO1}, we presented an alternative renormalization scheme for the
entanglement entropy (EE) of 3D conformal field theories (CFTs) with 4D
asymptotically anti-de Sitter (AAdS) gravity duals, in the context of the
gauge/gravity duality \cite{AdS/CFT1,AdS/CFT2,AdS/CFT3}. The scheme considers
the renormalized Einstein-AdS action obtained through the
\textit{Kounterterms} procedure
\cite{K1Even,K2Odd,K3AdS,KounterComparison,KounterComparison2}, evaluated on a
conically-singular manifold
\cite{Solodukhin1,SolodukhinNew,Solodukhin2,Cone3,Atiyah-Lebrun} defined via
the Replica Trick \cite{EECovariant,Maldacena1,Marika,RenyiXiDong}. The
renormalized EE thus obtained, corresponds to a modification of the RT area
functional \cite{RT1,TakayanagiReview}, which includes the addition of the
Chern form with a fixed coefficient. We now generalize this method to
holographic CFTs of arbitrary odd dimensions by considering the properties of
squashed-cones \cite{SolodukhinNew,Atiyah-Lebrun} and the renormalized
gravitational action given in Ref.\cite{K1Even}.

For clarity, we begin by reviewing the usual RT proposal
\cite{RT1,TakayanagiReview}, which considers that the EE of an entangling
region $A$ in a holographic CFT is given by the volume of a codimension-2
extremal surface $\Sigma$ in the AAdS bulk. This surface $\Sigma$ is such that
it has minimal area, its boundary $\partial\Sigma$ is located at the spacetime
boundary $B$, and $\partial\Sigma$ is conformal to the entangling surface
$\partial A$ which bounds $A$ at the conformal boundary $C$. We include a
pictorial representation of the different submanifolds involved in the RT
construction, explaining the relations between them, in FIG. 1.

\begin{figure}[h]
\begin{center}
\includegraphics[
height=6.0in,
width=6.0in
]{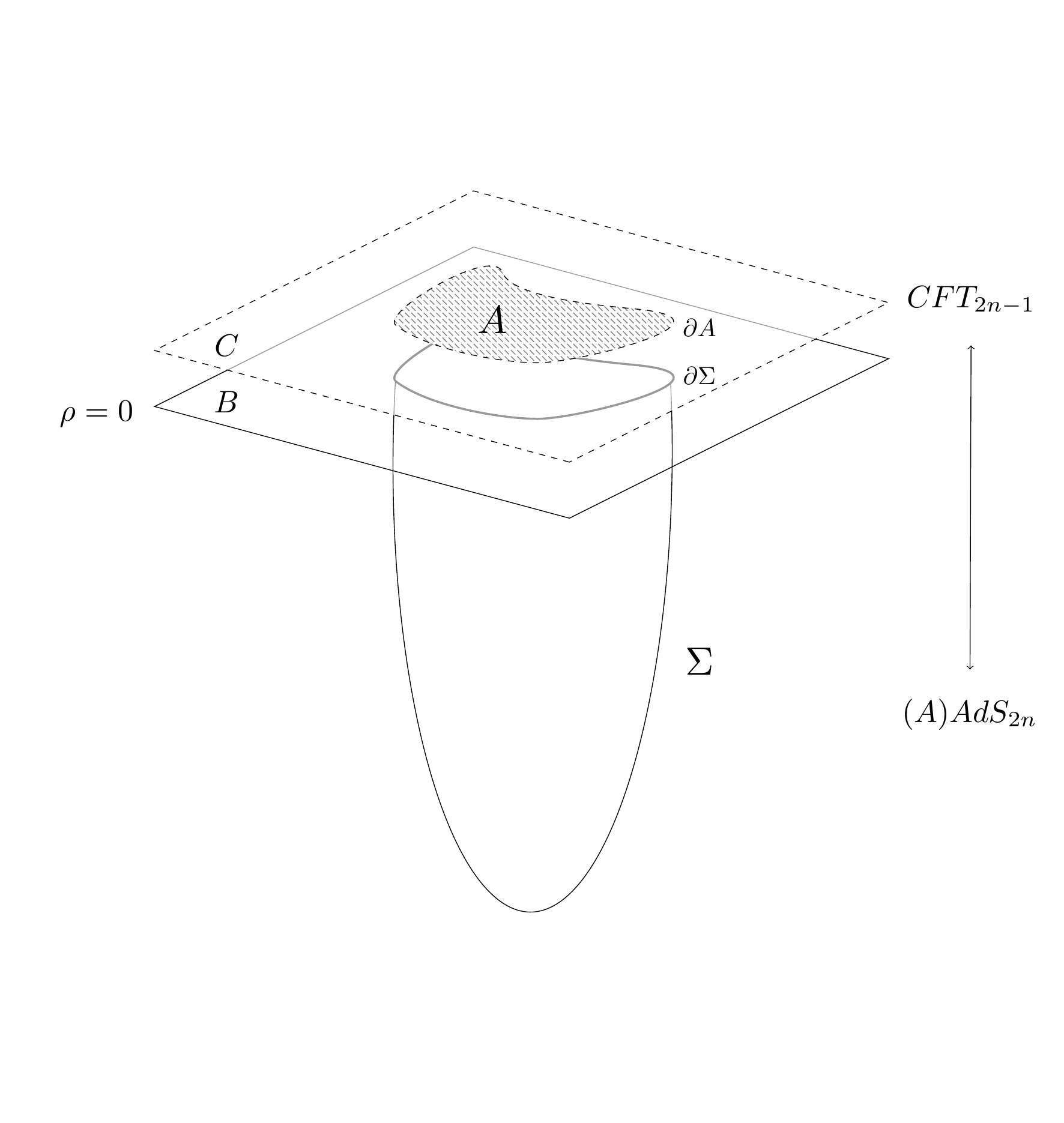}
\end{center}
\caption{In this diagram, we show all the submanifolds involved in the RT
construction. On the field theory side, $C$ is the conformal boundary where
the CFT is defined, $A$ is the entangling region and $\partial A$ is the
entangling surface. On the gravity side, $B$ is the boundary of spacetime,
$\Sigma$ is the minimal surface in the bulk and $\partial\Sigma$ is its border
at the spacetime boundary. Both sides are related such that $C$ is conformal
to $B$ and $\partial A$ is conformal to $\partial\Sigma$.}%
\end{figure}

As it is well known, the RT proposal for the EE is divergent due to the
presence of an infinite conformal factor at the boundary $B$. This is apparent
in the Fefferman-Graham (FG) expansion of the AAdS metric
\cite{FeffermanGraham,Imbimbo}. There are different methods for renormalizing
the EE. For example, the scheme proposed by Taylor and Woodhead \cite{Marika},
which is based on the standard Holographic Renormalization procedure
\cite{UsualCounterJohnson,DirichletKraus(quasiloc),DirichletHSS(quasiloc),
DirichletBalaKraus(quasiloc),DirichletHS(ano),SkenderisAndPapa1,SkenderisAndPapa2}%
. There is also the alternative topological renormalization scheme proposed in
Ref.\cite{AAO1}, which is based on the Kounterterms renormalization procedure
\cite{K1Even,K2Odd,K3AdS,KounterComparison,KounterComparison2}.

Both renormalization schemes rely on writing the EE in terms of the Euclidean
gravitational action in the bulk, including its corresponding counterterms at
the boundary $B$. As shown by Lewkowycz and Maldacena \cite{Maldacena1}, and
by Dong \cite{RenyiXiDong}, the replica trick can be used to construct a
suitable $2n-$dimensional bulk replica orbifold $\widehat{M}_{2n}^{(\alpha)}$,
which is a squashed-cone (conically singular manifold without U$\left(
1\right)  $ isometry) having conical angular parameter $\alpha$, such that
$2\pi\left(  1-\alpha\right)  $ is its angular deficit. Then, using the
AdS/CFT correspondence in the semi-classical limit, the EE can be expressed as%

\begin{equation}
S_{EE}=\left.  -\partial_{\alpha}I_{E}\left[  \widehat{M}_{2n}^{(\alpha
)}\right]  \right\vert _{\alpha=1}, \label{Replica_Alpha_EE}%
\end{equation}
where $I_{E}$ is the Euclidean gravitational action in the bulk, evaluated on
the $\widehat{M}_{2n}^{(\alpha)}$ orbifold. If $I_{E}$ is chosen as the
Euclidean Einstein-Hilbert (EH) action, then Eq.(\ref{Replica_Alpha_EE})
reproduces the RT proposal for the EE. However, it is apparent that if $I_{E}$
is chosen instead as a suitably renormalized gravitational action, then
$S_{EE}$ will be renormalized as well.

Here, we consider the renormalized Euclidean action $I_{E}^{ren}$ as given by
the Kounterterms scheme for even-dimensional manifolds in Ref.\cite{K1Even}.
Then we proceed in the same manner than in Ref.\cite{AAO1}, but now
considering that the action has to be evaluated on a squashed-cone of higher
even dimension (beyond the particular case of $D=4$). After taking the
corresponding derivative with respect to the angular parameter, we obtain that
the renormalized entanglement entropy $S_{EE}^{ren}$ is given by%

\begin{equation}
S_{EE}^{ren}=\frac{Vol\left(  \Sigma\right)  }{4G}+\frac{\left(  -1\right)
^{n}\ell^{2\left(  n-1\right)  }}{4G\left[  2\left(  n-1\right)  \right]
!}\int\limits_{\partial\Sigma}B_{2n-3}, \label{S_EE^ren}%
\end{equation}
where $\ell$ is the AdS radius, $Vol\left(  \Sigma\right)  $ is the volume of
the codimension-2 extremal surface $\Sigma$ and $B_{2n-3}$ is the $\left(
n-1\right)  -$th Chern form evaluated at $\partial\Sigma$, whose explicit form
is given in Eq.(\ref{(n-1)-thChern}). The derivation of Eq.(\ref{S_EE^ren})
and its properties are further explained in section \ref{Renorm}. The first
term of Eq.(\ref{S_EE^ren}) is exactly the RT proposal. Therefore, the
counterterm that renormalizes the EE is given by the $B_{2n-3}$ part, which
depends on the induced metric $\widetilde{\gamma}$ on $\partial\Sigma$, and on
both the intrinsic curvature of $\partial\Sigma$ and its extrinsic curvature
with respect to the radial foliation.

The expression for $S_{EE}^{ren}$ given in Eq.(\ref{S_EE^ren}) can be
rewritten in terms of the Euler characteristic and the AdS curvature
\cite{OleaF} of the minimal surface $\Sigma$, as explained in section
\ref{TopolInterp}. This is done by considering the usual Euler theorem for
regular manifolds, which relates the Chern form at $\partial\Sigma$ with the
Euler density on $\Sigma$. After some algebraic manipulations, $S_{EE}^{ren}$
can be expressed as%

\begin{gather}
S_{EE}^{ren}={\displaystyle\int\limits_{\Sigma}}\frac{d^{2n-2}y\sqrt{\gamma
}\ell^{2\left(  n-1\right)  }}{4G2^{\left(  n-1\right)  }\left(  2n-2\right)
!}{\displaystyle\sum\limits_{k=1}^{n-1}}\frac{\left(  -1\right)  ^{\left(
1+k\right)  }\left[  2\left(  n-1-k\right)  \right]  !2^{\left(  n-1-k\right)
}}{\ell^{2\left(  n-1-k\right)  }}\binom{n-1}{k}\times\nonumber\\
\times\delta_{\left[  a_{1}...a_{2k}\right]  }^{\left[  b_{1}...b_{2k}\right]
}\left(  \mathcal{F}_{\left.  AdS\right\vert _{\Sigma}}\right)  _{b_{1}b_{2}%
}^{a_{1}a_{2}}\cdots\left(  \mathcal{F}_{\left.  AdS\right\vert _{\Sigma}%
}\right)  _{b_{2k-1}b_{2k}}^{a_{2k-1}a_{2k}}+C_{\chi}%
\,,\label{TopolRenormIntro}%
\end{gather}
where%

\begin{equation}
C_{\chi}=\frac{\left(  -1\right)  ^{n+1}\left(  4\pi\right)  ^{\left(
n-1\right)  }\left(  n-1\right)  !\ell^{2\left(  n-1\right)  }}{4G\left(
2n-2\right)  !}\chi\left[  \Sigma\right]  \,,\nonumber
\end{equation}
and%

\begin{equation}
\left(  \mathcal{F}_{\left.  AdS\right\vert _{\Sigma}}\right)  _{b_{1}b_{2}%
}^{a_{1}a_{2}}=\mathcal{R}_{b_{1}b_{2}}^{a_{1}a_{2}}+\frac{1}{\ell^{2}}%
\delta_{\left[  b_{1}b_{2}\right]  }^{\left[  a_{1}a_{2}\right]  }\,,
\end{equation}
is the AdS curvature of $\Sigma$, which depends on its intrinsic Riemann
curvature tensor, and $\chi\left[  \Sigma\right]  $ is the Euler
characteristic of $\Sigma$. This topological form of $S_{EE}^{ren}$ is useful
for computing the renormalized EE of certain entangling regions, which give
rise to constant curvature minimal surfaces in the bulk. For example, in
section \ref{Explicit}, we discuss the case of ball-shaped entangling regions,
where the computation of $S_{EE}^{ren}$ is greatly simplified.

The organization of this paper is as follows: In section \ref{Euler}, we
discuss a possible generalization of the Euler theorem for squashed cones in
even dimensions. In section \ref{Renorm}, we derive $S_{EE}^{ren}$ by
introducing the renormalized gravitational action for AAdS$_{2n}$ evaluated on
the replica orbifold. We verify the cancellation of the divergence in
$S_{EE}^{ren}$ up to next-to-leading order in the radial coordinate $\rho$. We
also reinterpret $S_{EE}^{ren}$ in terms of the topological and geometric
properties of $\Sigma$. In section \ref{Explicit}, we rederive $S_{EE}^{ren}$
for spherical entangling surfaces in odd-dimensional CFTs by using the
topological renormalization scheme of Eq.(\ref{TopolRenormIntro}), recovering
the results of Refs.\cite{NishiokaPaper,NishiokaReview}. We also comment on
the simplicity of the computation using the topological scheme. Finally, in
section \ref{Outlook}, we comment on the application of the renormalization
procedure for characterizing odd-dimensional CFTs and future generalizations thereof.

\section{Euler density for even-dimensional squashed cones\label{Euler}}

In Ref.\cite{AAO1}, we discussed the Euler theorem for squashed cones in 4D,
which was derived by Fursaev, Patrushev and Solodukhin in
Ref.\cite{SolodukhinNew} considering the form of the quadratic curvature
invariants on squashed-cone manifolds, which have non-trivial extrinsic
curvature contributions coming from the binormal foliation at the tip of the
cone. The theorem states that%

\begin{equation}
\int\limits_{M_{4}^{\left(  \alpha\right)  }}\varepsilon_{4}^{\left(
\alpha\right)  }=\int\limits_{M_{4}}\varepsilon_{4}^{\left(  r\right)  }%
+8\pi\left(  1-\alpha\right)  \int\limits_{\Sigma}\varepsilon_{2}+O\left(
\left(  1-\alpha\right)  ^{2}\right)  , \label{Euler_conic_4D}%
\end{equation}
where $\varepsilon_{4}^{\left(  \alpha\right)  }$ is the Euler density
evaluated on the 4D squashed-cone manifold $\widehat{M}_{4}^{\left(
\alpha\right)  }$, $\varepsilon_{4}^{\left(  r\right)  }$ is the Euler density
evaluated on a regular manifold $M_{4}$ given by the $\alpha\rightarrow1$
limit and $\varepsilon_{2}$ is the Euler density evaluated on the 2D manifold
$\Sigma$, which is the codimension-2 surface located at the tip of the cone
(defined for integer $\left(  \frac{1}{\alpha}\right)  $ as the fixed-point
set of the $Z_{\left(  \frac{1}{\alpha}\right)  }$ symmetry). The form of
$\varepsilon_{4}$, which is also referred to as the Gauss-Bonnet term, is
given in terms of the quadratic curvature invariants by%

\begin{equation}
\varepsilon_{4}=\sqrt{G}d^{4}x\left(  R_{\nu\sigma\lambda}^{\mu}R_{\mu}%
^{\nu\sigma\lambda}-4R_{\mu\nu}R^{\mu\nu}+R^{2}\right)  ,
\end{equation}
and $\varepsilon_{2}=\mathcal{R}\sqrt{\gamma}d^{2}x$, where the cursive
$\mathcal{R}$ denotes the Ricci scalar on $\Sigma$.

Furthermore in Ref.\cite{AAO1}, we derived a relation between the second Chern
form at the spacetime boundary and the first Chern form at $\partial\Sigma$,
given by

\bigskip%
\begin{equation}
\int\limits_{\partial\widehat{M}_{4}^{\left(  \alpha\right)  }}B_{3}^{\left(
\alpha\right)  }=\int\limits_{\partial M_{4}}B_{3}^{\left(  r\right)  }%
+8\pi\left(  1-\alpha\right)  \int\limits_{\partial\Sigma}B_{1}+O\left(
\left(  1-\alpha\right)  ^{2}\right)  . \label{Euler_Boundary_Conical_4D}%
\end{equation}
In order to obtain this, we considered the usual Euler theorem for regular
$2n-$dimensional manifolds $M_{2n}$, which states that%

\begin{equation}
\int\limits_{M_{2n}}\varepsilon_{2n}=\left(  4\pi\right)  ^{n}n!\chi\left[
M_{2n}\right]  +\int\limits_{\partial M_{2n}}B_{2n-1}, \label{Usual_Euler}%
\end{equation}
where $B_{2n-1}$ is the $n-$th Chern form at the boundary $\partial M_{2n}$
(whose form in Gauss normal coordinates is given in Eq.(\ref{B_2n-1})) and
$\chi\left[  X\right]  $ is the Euler characteristic of the manifold $X$. We
also considered the relation satisfied by the Euler characteristic of 4D
squashed-cones, given by%

\begin{equation}
\chi\left[  \widehat{M}_{4}^{\left(  \alpha\right)  }\right]  =\chi\left[
M_{4}\right]  +\left(  1-\alpha\right)  \chi\left[  \Sigma\right]  +O\left(
\left(  1-\alpha\right)  ^{2}\right)  ,\label{Euler_Char_Cone}%
\end{equation}
and derived by FPS in Ref.\cite{SolodukhinNew}. Then,
Eq.(\ref{Euler_Boundary_Conical_4D}) can be obtained by considering $n=1$ and
$n=2$ in Eq.(\ref{Usual_Euler}), replacing the $\varepsilon_{4}$ for the
$B_{3}$ and the $\varepsilon_{2}$ for the $B_{1}$ in Eq.(\ref{Euler_conic_4D}%
), and then using Eq.(\ref{Euler_Char_Cone}) to eliminate the Euler characteristics.

We now conjecture that Eqs.(\ref{Euler_conic_4D}%
,\ref{Euler_Boundary_Conical_4D},\ref{Euler_Char_Cone}) have generalizations
for squashed-cone manifolds of arbitrary even dimensions. Namely, we propose that%

\begin{equation}
\int\limits_{\widehat{M}_{2n}^{\left(  \alpha\right)  }}\varepsilon
_{2n}^{\left(  \alpha\right)  }=\int\limits_{M_{2n}}\varepsilon_{2n}^{\left(
r\right)  }+4n\pi\left(  1-\alpha\right)  \int\limits_{\Sigma}\varepsilon
_{2\left(  n-1\right)  }+O\left(  \left(  1-\alpha\right)  ^{2}\right)  ,
\label{Euler_Conic}%
\end{equation}

\begin{equation}
\int\limits_{\partial\widehat{M}_{2n}^{\left(  \alpha\right)  }}%
B_{2n-1}^{\left(  \alpha\right)  }=\int\limits_{\partial M_{2n}}%
B_{2n-1}^{\left(  r\right)  }+4n\pi\left(  1-\alpha\right)  \int
\limits_{\partial\Sigma}B_{2n-3}+O\left(  \left(  1-\alpha\right)
^{2}\right)  \label{Euler_Conic_Boundary}%
\end{equation}
and%

\begin{equation}
\chi\left[  \widehat{M}_{2n}^{\left(  \alpha\right)  }\right]  =\chi\left[
M_{2n}\right]  +\left(  1-\alpha\right)  \chi\left[  \Sigma\right]  +O\left(
\left(  1-\alpha\right)  ^{2}\right)  . \label{Euler_Char_2n}%
\end{equation}
In the above relations, the Euler density $\varepsilon_{2n}$ is given by%

\begin{equation}
\varepsilon_{2n}=\frac{1}{2^{n}}d^{2n}x\sqrt{G}\delta_{\left[  \mu_{1}%
...\mu_{2n}\right]  }^{\left[  \nu_{1}...\nu_{2n}\right]  }R_{\nu_{1}\nu_{2}%
}^{\mu_{1}\mu_{2}}...R_{\nu_{2n-1}\nu_{2n}}^{\mu_{2n-1}\mu_{2n}},
\label{Euler_2n}%
\end{equation}
$B_{2n-1}^{\left(  \alpha\right)  }$ is the $n-$th Chern form evaluated at the
conically-singular $\partial\widehat{M}_{2n}^{\left(  \alpha\right)  }$
boundary, $B_{2n-1}^{\left(  r\right)  }$ is the $n-$th Chern form evaluated
at the regular $\partial M_{2n}$ boundary (corresponding to the $\alpha
\rightarrow1$ limit) and $B_{2n-3}$ is the $\left(  n-1\right)  -$th Chern
form evaluated at $\partial\Sigma$. Also, $\delta_{\left[  \mu_{1}...\mu
_{2n}\right]  }^{\left[  \nu_{1}...\nu_{2n}\right]  }$ is the totally
antisymmetric generalization of the Kronecker delta, defined by
\begin{equation}
\delta_{\left[  \mu_{1}...\mu_{2n}\right]  }^{\left[  \nu_{1}...\nu
_{2n}\right]  }\overset{def}{=}\det\left[  \delta_{\mu_{1}}^{\nu_{1}}%
\cdots\delta_{\mu_{k}}^{\nu_{k}}\right]  .
\end{equation}

We mention that Eqs.(\ref{Euler_Conic}-\ref{Euler_Char_2n}) were proven by
Fursaev and Solodukhin in Ref.\cite{Solodukhin1} for the case of
conically-singular manifolds with a U$\left(  1\right)  $ rotational isometry
about the symmetry axis of the cone. Here we simply assume these equations to
hold also for the squashed-cone case, and we show that if they are correct,
then the $S_{EE}^{ren}$ for arbitrary odd-dimensional holographic CFTs can be
obtained in a manner analogous to the 3D case studied in Ref.\cite{AAO1}.
Conversely, as we will explicitly verify (up to next-to-leading order in the
holographic radial coordinate $\rho$) in section \ref{Finiteness}, the
$S_{EE}^{ren}$ obtained in this manner does indeed renormalize the EE, which
lends credence to our conjectured generalization of the Euler theorem,
although it does not constitute a proof thereof.

The expression given in Eq.(\ref{Euler_Conic_Boundary}) is precisely what is
used in section \ref{Renorm} in order to evaluate the Euclidean action in the
replica orbifold, as required in the computation of $S_{EE}$ according to
Eq.(\ref{Replica_Alpha_EE}). This ultimately gives the expression for
$S_{EE}^{ren}$ when considering the renormalized Euclidean action for
AAdS$_{2n}$, which is discussed in the following section.

\section{Renormalization of EE in AdS$_{2n}$/CFT$_{2n-1}$ through the Chern
form\label{Renorm}}

In this section, we study the renormalization of EE via the topological
scheme. We consider the renormalized Euclidean gravity action for AAdS$_{2n}$
spacetimes given in Ref.\cite{K1Even}, which was obtained by the Kounterterms
procedure \cite{K1Even,K2Odd,K3AdS,KounterComparison,KounterComparison2}. The
preference for this renormalization procedure over the usual holographic
renormalization scheme is only due to practical reasons, as both schemes have
been shown to give assymptotically equivalent results for the renormalized
action \cite{KounterComparison,KounterComparison2}. Essentially, as explained
in Ref.\cite{AAO1}, the renormalization of the even-dimensional action in the
Kounterterms scheme can be accomplished by the addition of the Chern form
(with a specific coupling), which constitutes a single boundary counterterm,
whereas in the case of holographic renormalization, the number of required
counterterms rapidly grows with the dimension.

The renormalized bulk action, when evaluated in the replica orbifold (as
described in the introduction), is given by%

\begin{equation}
I_{E}^{ren}=\frac{1}{16\pi G}%
{\displaystyle\int\limits_{\hat{M}_{2n}^{\left(  \alpha\right)  }}}
d^{2n}x\sqrt{G}\left(  R^{\left(  \alpha\right)  }-2\Lambda\right)
+\frac{c_{2n}}{16\pi G}%
{\displaystyle\int\limits_{\partial\hat{M}_{2n}^{\left(  \alpha\right)  }}}
B_{2n-1}^{\left(  \alpha\right)  }, \label{I_ren_alpha}%
\end{equation}
where%

\begin{equation}%
\begin{tabular}
[c]{l}%
$\Lambda=-\frac{\left(  2n-1\right)  \left(  2n-2\right)  }{2\ell^{2}},$\\
$c_{2n}=\frac{\left(  -1\right)  ^{n}\ell^{2\left(  n-1\right)  }}{n\left[
2\left(  n-1\right)  \right]  !}$%
\end{tabular}
\ \ \label{action_constants}%
\end{equation}
and the $n-$th Chern form $B_{2n-1}$ is given by%

\begin{gather}
B_{2n-1}=-2n{\displaystyle\int\limits_{0}^{1}}dtd^{2n-1}x\sqrt{h}%
\delta_{\lbrack i_{1}...i_{2n-1}]}^{[j_{1}...j_{2n-1}]}K_{j_{1}}^{i_{1}%
}\left(  \frac{1}{2}\mathcal{R}_{j_{2}j_{3}}^{i_{2}i_{3}}-t^{2}K_{j_{2}%
}^{i_{2}}K_{j_{3}}^{i_{3}}\right)  \times\ldots\nonumber\\
\ldots\times\left(  \frac{1}{2}\mathcal{R}_{j_{2n-2}j_{2n-1}}^{i_{2n-2}%
i_{2n-1}}-t^{2}K_{j_{2n-2}}^{i_{2n-2}}K_{j_{2n-1}}^{i_{2n-1}}\right)
\,.\label{B_2n-1}%
\end{gather}
In the expression for the Chern form of Eq.(\ref{B_2n-1}), $h_{ij}$ is the
metric at the spacetime boundary $B$, $\mathcal{R}_{ij}^{k\ell}$ is the
intrinsic Riemann tensor computed with $h_{ij}$, and $K_{j}^{i}$ is the
extrinsic curvature of $B$ with respect to the radial foliation. Also, we note
that in Eq.(\ref{I_ren_alpha}), $B_{2n-1}^{\left(  \alpha\right)  }$ denotes
the Chern form evaluated at the conically singular boundary of the $\hat
{M}_{2n}^{\left(  \alpha\right)  }$ orbifold, which can be expressed in terms
of the Chern form at the boundary of the minimal surface $\Sigma$ using our
conjectured generalization of the Euler theorem to squashed-cones in arbitrary
even dimensions, as presented in Eq.(\ref{Euler_Conic_Boundary}).

When computing the EE, after taking the $\alpha$ derivative according to
Eq.(\ref{Replica_Alpha_EE}), the Einstein-Hilbert part simply gives the usual
RT minimal area prescription for the EE, as shown by Lewkowicz and Maldacena
\cite{Maldacena1} and by Dong \cite{RenyiXiDong}. The EE counterterm then
comes from the term containing the boundary Chern form $B_{2n-1}^{\left(
\alpha\right)  }$. We therefore define the counterterm of the Euclidean action as%

\begin{equation}
I_{E}^{ct}=\frac{c_{2n}}{16\pi G}%
{\displaystyle\int\limits_{\partial\hat{M}_{2n}^{\left(  \alpha\right)  }}}
B_{2n-1}^{\left(  \alpha\right)  }, \label{I_ct}%
\end{equation}
and we proceed to compute the counterterm of the EE ($S_{EE}^{ct}$) as%

\begin{equation}
S_{EE}^{ct}=\left.  -\partial_{\alpha}I_{E}^{ct}\left(  \partial\widehat
{M}_{2n}^{(\alpha)}\right)  \right\vert _{\alpha=1}, \label{S_ct}%
\end{equation}
such that $S_{EE}^{ren}=S_{EE}^{RT}+S_{EE}^{ct}$, where $S_{EE}^{RT}$ is the
usual RT prescription for the EE. Finally, using
Eq.(\ref{Euler_Conic_Boundary}) to evaluate $I_{E}^{ct}$, we obtain%

\begin{equation}
S_{EE}^{ct}=\frac{\left(  -1\right)  ^{n}\ell^{2\left(  n-1\right)  }%
}{4G\left[  2\left(  n-1\right)  \right]  !}\int\limits_{\partial\Sigma
}B_{2n-3}, \label{S_EE^ct}%
\end{equation}
recovering $S_{EE}^{ren}$ as given in Eq.(\ref{S_EE^ren}).

\subsection{Explicit covariant embedding}

Now, we consider the embedding of the minimal surface $\Sigma$ in the AAdS
bulk as given by Hung, Myers and Smolkin \cite{HungMyersSmolkin}, and by
Schwimmer and Theisen \cite{SchwimmerAndTheisen}. The embedding is such that
the bulk coordinates $\left\{  \rho,x^{i}\right\}  $ of $\Sigma$ and its
worldvolume coordinates $\left\{  \tau,y^{a}\right\}  $ are related by%

\begin{equation}%
\begin{tabular}
[c]{l}%
$x^{i}\left(  \tau,y^{a}\right)  =\left(  x^{\left(  0\right)  }\right)
^{i}\left(  y^{a}\right)  +\tau\left(  x^{\left(  2\right)  }\right)
^{i}\left(  y^{a}\right)  +...$\\
$\left(  x^{\left(  2\right)  }\right)  ^{i}\left(  y^{a}\right)  =\frac
{\ell^{2}}{2\left(  d-2\right)  }\kappa^{i}\left(  y^{a}\right)  .$%
\end{tabular}
\ \ \ \label{Sigma_embedding}%
\end{equation}

We also consider the asymptotic expansion of the bulk metric $G_{\mu\nu}$,
which is of FG form \cite{FeffermanGraham} and is given by%

\begin{equation}%
\begin{tabular}
[c]{l}%
$ds_{G}^{2}=G_{\mu\nu}dx^{\mu}dx^{\nu}=\frac{\ell^{2}d\rho^{2}}{4\rho^{2}%
}+h_{ij}\left(  \rho,x\right)  dx^{i}dx^{j},$\\
$h_{ij}\left(  \rho,x\right)  =\frac{g_{ij}\left(  \rho,x\right)  }{\rho},$\\
$g_{ij}\left(  \rho,x\right)  =g_{ij}^{\left(  0\right)  }\left(  x\right)
+\rho g_{ij}^{\left(  2\right)  }\left(  x\right)  +...,$%
\end{tabular}
\ \ \ \ \ \label{F-G_BulkMetric}%
\end{equation}
as presented in Ref.\cite{Imbimbo}. Here, $\rho$ is the holographic radial
coordinate and $h_{ij}$ is the induced metric at the spacetime boundary $B$,
which is located at $\rho=0$.

We now consider the induced metric $\gamma_{ab}$ at the minimal surface
$\Sigma$, which is defined as the pullback of $G_{\mu\nu}$ in the FG gauge,
and is therefore given by%

\begin{equation}
\gamma_{ab}=\frac{\partial x^{\mu}}{\partial y^{a}}\frac{\partial x^{\nu}%
}{\partial y^{b}}G_{\mu\nu}. \label{InducedMetricSigma}%
\end{equation}
Fixing the diffeomorphism gauge as $\tau=\rho$ and $\gamma_{a\tau}=0$, we
obtain a FG-like expansion for $\gamma_{ab}$ such that%

\begin{equation}%
\begin{tabular}
[c]{l}%
$ds_{\gamma}^{2}=\gamma_{ab}dy^{a}dy^{b}=\frac{\ell^{2}}{4\tau^{2}}\left(
1+\frac{\tau\ell^{2}}{\left(  d-2\right)  ^{2}}\kappa^{i}\kappa^{j}%
g_{ij}^{\left(  0\right)  }+...\right)  d\tau^{2}+\widetilde{\gamma}%
_{ab}\left(  \tau,y\right)  dy^{a}dy^{b},$\\
$\widetilde{\gamma}_{ab}\left(  \tau,y\right)  =\frac{\sigma_{ab}\left(
\tau,y\right)  }{\tau},$\\
$\sigma_{ab}\left(  \tau,y\right)  =\sigma_{ab}^{\left(  0\right)  }\left(
y\right)  +\tau\sigma_{ab}^{\left(  2\right)  }\left(  y\right)  +...~.$%
\end{tabular}
\ \ \ \ \ \label{F-G_SigmaMetric}%
\end{equation}

In the previous expansions, $d$ is defined as the dimension of the spacetime
boundary ($d=2n-1$). Also, $g_{ij}^{\left(  0\right)  }$ corresponds to the
metric of the CFT at the conformal boundary $C$ (conformal to $B$) and
$\sigma_{ab}^{\left(  0\right)  }$ denotes the induced metric on the
entangling surface $\partial A$ (conformal to $\partial\Sigma$) which is given
by
\begin{equation}
\sigma_{ab}^{\left(  0\right)  }=\frac{\partial\left(  x^{\left(  0\right)
}\right)  ^{i}}{\partial y^{a}}\frac{\partial\left(  x^{\left(  0\right)
}\right)  ^{j}}{\partial y^{b}}g_{ij}^{\left(  0\right)  }.
\end{equation}
Moreover, $g_{ij}^{\left(  2\right)  }=-\ell^{2}S_{ij}^{\left(  0\right)  }$
where $S_{ij}^{\left(  0\right)  }$, defined as
\begin{equation}
\left(  S^{\left(  0\right)  }\right)  _{ij}=\frac{1}{\left(  2n-3\right)
}\left(  R_{ij}^{\left(  0\right)  }-\frac{g_{ij}^{\left(  0\right)  }%
}{2\left(  2n-2\right)  }R^{\left(  0\right)  }\right)  ,
\end{equation}
denotes the Schouten tensor of the $g_{ij}^{\left(  0\right)  }$. Now,
considering the definition of $\gamma_{ab}$ given in
Eq.(\ref{InducedMetricSigma}) and the embedding of $\Sigma$ from
Eq.(\ref{Sigma_embedding}), we obtain that
\begin{equation}
\sigma_{ab}^{\left(  2\right)  }=\frac{\partial\left(  x^{\left(  0\right)
}\right)  ^{i}}{\partial y^{a}}\frac{\partial\left(  x^{\left(  0\right)
}\right)  ^{j}}{\partial y^{b}}g_{ij}^{\left(  2\right)  }-\frac{\ell^{2}%
}{\left(  2n-3\right)  }\kappa^{i}\kappa_{ab}^{j}\left(  g^{\left(  0\right)
}\right)  _{ij}. \label{Sigma_2}%
\end{equation}
Here,
\begin{equation}
\kappa^{i}=\hat{n}_{\left(  n\right)  }^{i}\kappa_{ab}^{\left(  n\right)
}\left(  \sigma^{\left(  0\right)  }\right)  ^{ab},
\end{equation}
such that $\hat{n}_{\left(  n\right)  }^{i}$ are the orthonormal vectors to
$\partial A$ at the conformal boundary $C$ ($n=1,2$), and $\kappa
_{ab}^{\left(  n\right)  }$ are the corresponding extrinsic curvatures of
$\partial A$.

In the following subsection, given the explicit embedding of $\Sigma$ in
$\widehat{M}_{2n}^{(\alpha)}$ and the corresponding FG expansions of
$G_{\mu\nu}$ and $\gamma_{ab}$, we proceed to verify the finiteness of
$S_{EE}^{ren}$, as given in Eq.(\ref{S_EE^ren}).

\subsection{Proof of finiteness of S$_{EE}^{ren}$\label{Finiteness}}

We now use the previously discussed embedding of $\Sigma$ in order to verify
that $S_{EE}^{ren}$, as defined in Eq.(\ref{S_EE^ren}), is free from
divergences. In order to do this, we first exhibit the divergence structure of
$S_{EE}^{RT}$, and then we check that the divergences are exactly cancelled by
the $S_{EE}^{ct}$ defined in Eq.(\ref{S_EE^ct}), without modifying the finite
universal part. We mention that although the value of $S_{EE}^{ren}$ depends
on the particular choice of entangling surface $\partial A$, the divergence
structure of $S_{EE}^{RT}$ and $S_{EE}^{ct}$ do not.

We start by computing the RT part of the EE, according to the minimal area
prescription \cite{RT1,TakayanagiReview}. We have that%

\begin{equation}
S_{EE}^{RT}=\frac{1}{4G}\int\limits_{\Sigma}d^{2n-2}y\sqrt{\gamma}=\frac
{1}{4G}\int\limits_{\partial\Sigma_{\varepsilon}}d^{2n-3}y%
{\displaystyle\int\limits_{\varepsilon}^{\rho_{\text{max}}}}
d\rho\sqrt{\gamma}, \label{S_RT}%
\end{equation}
where $\rho_{\text{max}}$ is the maximum value of the holographic radial
coordinate on the $\Sigma$ surface, which depends on the choice of entangling
surface $\partial A$ in the CFT, and $\varepsilon$ is a cutoff such that the
$\varepsilon\rightarrow0$ limit is to be evaluated at the end. Considering the
FG-like expansion of $\gamma_{ab}$ given in Eq.(\ref{F-G_SigmaMetric}), we
have that%

\begin{equation}
\sqrt{\gamma}=\frac{\ell\sqrt{\sigma^{\left(  0\right)  }}}{2\rho^{\left(
2n-1\right)  /2}}\left(  1+\rho\left[  \frac{\ell^{2}}{2\left(  2n-3\right)
^{2}}\kappa^{i}\kappa^{j}g_{ij}^{\left(  0\right)  }+\frac{1}{2}%
tr[\sigma^{\left(  2\right)  }]\right]  +...\right)  ,
\end{equation}
and therefore, by performing the $\rho$ integration, we obtain%

\begin{equation}%
\begin{tabular}
[c]{l}%
$%
{\displaystyle\int\limits_{\varepsilon}^{\rho_{\text{max}}}}
d\rho\sqrt{\gamma}=C_{1}+\frac{\ell\sqrt{\sigma^{\left(  0\right)  }}}{\left(
2n-3\right)  \varepsilon^{\left(  2n-3\right)  /2}}\left(  1+\varepsilon
\left[  \frac{\left(  2n-3\right)  }{2\left(  2n-5\right)  }tr[\sigma^{\left(
2\right)  }]+\frac{\ell^{2}}{2\left(  2n-5\right)  \left(  2n-3\right)
}\kappa^{i}\kappa^{j}g_{ij}^{\left(  0\right)  }\right]  +...\right)  ,$%
\end{tabular}
\ \ \label{Rho_Int}%
\end{equation}
where $C_{1}$ is a finite constant that depends on the value of $\rho
_{\text{max}}$. Here we can see that the leading and next-to-leading
divergences occur at orders $\varepsilon^{-\frac{\left(  2n-3\right)  }{2}}$
and $\varepsilon^{-\frac{\left(  2n-5\right)  }{2}}$respectively, as expected.

Now, we consider the form of $\sigma_{ab}^{\left(  2\right)  }$, which is the
second coefficient in the FG expansion of the induced metric $\widetilde
{\gamma}_{ab}$ at $\partial\Sigma$ and it is given in Eq.(\ref{Sigma_2}). We
therefore have that%

\begin{equation}
tr\left[  \sigma^{\left(  2\right)  }\right]  =-\ell^{2}S\left[
\sigma^{\left(  0\right)  }\right]  -\frac{\ell^{2}}{2\left(  2n-3\right)
}\kappa^{i}\kappa^{j}g_{ij}^{\left(  0\right)  }, \label{Trace-of-sigma2}%
\end{equation}
where $S\left[  \sigma^{\left(  0\right)  }\right]  $ is the trace of the
Schouten tensor of the induced metric $\sigma_{ab}^{\left(  0\right)  }$ at
$\partial A$. Also, considering the definition of the Schouten tensor, its
trace can be related to the Ricci scalar of $\sigma_{ab}^{\left(  0\right)  }$ as%

\begin{equation}
S\left[  \sigma^{\left(  0\right)  }\right]  =\frac{\mathcal{R}^{\left(
0\right)  }}{2\left(  2n-4\right)  }.
\end{equation}
Finally, we obtain%

\begin{equation}
tr\left[  \sigma^{\left(  2\right)  }\right]  =-\ell^{2}\left(  \frac
{\mathcal{R}^{\left(  0\right)  }}{2\left(  2n-4\right)  }+\frac{\kappa
^{i}\kappa^{j}g_{ij}^{\left(  0\right)  }}{2\left(  2n-3\right)  }\right)  .
\label{Tr_Sigma2}%
\end{equation}

By replacing Eq.(\ref{Tr_Sigma2}) into the radial integral of
Eq.(\ref{Rho_Int}), and after some simplifications, we can rewrite
$S_{EE}^{RT}$ as%

\begin{equation}%
\begin{tabular}
[c]{l}%
$S_{EE}^{RT}=\frac{C_{2}}{4G}+\frac{1}{4G}%
{\displaystyle\int\limits_{\partial\Sigma_{\varepsilon}}}
d^{2n-3}y\frac{\ell\sqrt{\sigma^{\left(  0\right)  }}}{\left(  2n-3\right)
\varepsilon^{\left(  2n-3\right)  /2}}-$\\
$\frac{1}{4G}%
{\displaystyle\int\limits_{\partial\Sigma_{\varepsilon}}}
d^{2n-3}y\frac{\ell^{3}\sqrt{\sigma^{\left(  0\right)  }}}{2\left(
2n-3\right)  ^{2}\left(  2n-5\right)  \varepsilon^{\left(  2n-5\right)  /2}%
}\left(  \left(  \frac{2n-5}{2}\right)  \kappa^{i}\kappa^{j}g_{ij}^{\left(
0\right)  }+\frac{\left(  2n-3\right)  ^{2}}{2\left(  2n-4\right)
}\mathcal{R}^{\left(  0\right)  }\right)  +...,$%
\end{tabular}
\ \ \ \label{S_EE^RT}%
\end{equation}
where $\frac{C_{2}}{4G}$ is the finite part of the EE, which depends on the
choice of $\partial A$.

We now analyze the asymptotic behavior of the EE counterterm, according to
Eq.(\ref{S_EE^ct}). We have that%

\begin{gather}
B_{2n-3}=-2\left(  n-1\right)  {\displaystyle\int\limits_{0}^{1}}%
dtd^{2n-3}y\sqrt{\widetilde{\gamma}}\delta_{\lbrack b_{1}...b_{2n-3}]}%
^{[a_{1}...a_{2n-3}]}k_{b_{1}}^{a_{1}}\left(  \frac{1}{2}\mathcal{R}%
_{b_{2}b_{3}}^{a_{2}a_{3}}-t^{2}k_{b_{2}}^{a_{2}}k_{b_{3}}^{a_{3}}\right)
\times\ldots\nonumber\\
\ldots\times\left(  \frac{1}{2}\mathcal{R}_{b_{2n-4}b_{2n-3}}^{a_{2n-4}%
a_{2n-3}}-t^{2}k_{b_{2n-4}}^{a_{2n-4}}k_{b_{2n-3}}^{a_{2n-3}}\right)
\,,\label{(n-1)-thChern}%
\end{gather}
where $\mathcal{R}_{b_{1}b_{2}}^{a_{1}a_{2}}$ is the Riemann tensor of the
$\widetilde{\gamma}$ metric at $\partial\Sigma$, $k_{b}^{a}=\widetilde{\gamma
}^{ac}k_{cb}$ and $k_{cb}$ is the extrinsic curvature of $\partial\Sigma$ with
respect to the radial foliation along the holographic coordinate $\rho$ (not
to be confused with $\kappa_{ab}^{\left(  n\right)  }$ for $\partial A$ or
with $K_{ij}$ for $B$). By definition,%

\begin{equation}
k_{ab}=\frac{-1}{2\sqrt{\gamma_{\rho\rho}}}\partial_{\rho}\widetilde{\gamma
}_{ab},
\end{equation}
and therefore, using the FG expansion for $\gamma_{ab}$ given in
Eq.(\ref{F-G_SigmaMetric}) and considering that%

\begin{equation}
\widetilde{\gamma}^{ab}=\rho\left(  \left(  \sigma^{\left(  0\right)
}\right)  ^{ab}-\rho\left(  \sigma^{\left(  2\right)  }\right)  ^{ab}\right)
,
\end{equation}
we have that%

\begin{equation}
k_{b}^{a}=\frac{1}{\ell}\left(  \delta_{b}^{a}-\rho\left[  \left(
\sigma^{\left(  2\right)  }\right)  _{b}^{a}+\frac{\ell^{2}\kappa^{i}%
\kappa^{j}g_{ij}^{\left(  0\right)  }}{2\left(  2n-3\right)  ^{2}}\delta
_{b}^{a}\right]  +...\right)  . \label{k-expansion}%
\end{equation}
We also have that $\sqrt{\widetilde{\gamma}}$ is given by%

\begin{equation}
\sqrt{\widetilde{\gamma}}=\frac{\sqrt{\sigma^{\left(  0\right)  }}}%
{\rho^{\left(  2n-3\right)  /2}}\left(  1+\frac{\rho}{2}tr[\sigma^{\left(
2\right)  }]+...\right)  , \label{sqrt_gamma_tilde}%
\end{equation}
and that the Riemann tensor of $\widetilde{\gamma}_{ab}$ satisfies
\begin{equation}
\mathcal{R}_{b_{1}b_{2}}^{a_{1}a_{2}}=\left(  \rho\left(  \mathcal{R}^{\left(
0\right)  }\right)  _{b_{1}b_{2}}^{a_{1}a_{2}}+...\right)  ,
\label{Riemann_Gamma}%
\end{equation}
where $\left(  \mathcal{R}^{\left(  0\right)  }\right)  _{b_{1}b_{2}}%
^{a_{1}a_{2}}$ is the Riemann tensor of the $\sigma_{ab}^{\left(  0\right)  }$
metric. We now have everything we need in order to expand $B_{2n-3}$ to
next-to-leading order in $\rho$. The explicit step-by-step computation is
presented in appendix \ref{AppA}.

We therefore obtain that the $\left(  n-1\right)  -$th Chern form at
$\partial\Sigma$, located at $\rho=\varepsilon$, is given by%

\begin{align}
B_{2n-3} &  =\frac{2\left(  n-1\right)  \left(  -1\right)  ^{\left(
n-1\right)  }\left[  2\left(  n-2\right)  \right]  !}{\ell^{\left(
2n-3\right)  }}\frac{d^{2n-3}y\sqrt{\sigma_{\alpha\beta}^{\left(  0\right)  }%
}}{\varepsilon^{\left(  2n-3\right)  /2}}\left\{  1-\frac{\varepsilon
}{2\left(  2n-3\right)  \left(  2n-5\right)  }\times\right.  \nonumber\\
&  \times\left.  \left[  \left(  2n-3\right)  \left(  2n-5\right)
tr[\sigma^{\left(  2\right)  }]+\left(  2n-5\right)  \kappa^{i}\kappa
^{j}g_{ij}^{\left(  0\right)  }\ell^{2}+\left(  2n-3\right)  \ell
^{2}\mathcal{R}^{\left(  0\right)  }\right]  +\cdots\right\}
\,,\label{B2n-3MainText}%
\end{align}
and thus, $S_{EE}^{ct}$ is given by%

\begin{gather}
S_{EE}^{ct}=-\frac{\ell}{4G\left(  2n-3\right)  }{\displaystyle\int
\limits_{\partial\Sigma_{\varepsilon}}}d^{2n-3}y\frac{\sqrt{\sigma
_{\alpha\beta}^{\left(  0\right)  }}}{\varepsilon^{\left(  2n-3\right)  /2}%
}\left\{  1-\frac{\varepsilon}{2\left(  2n-3\right)  \left(  2n-5\right)
}\left[  \left(  2n-3\right)  \left(  2n-5\right)  tr[\sigma^{\left(
2\right)  }]+\right.  \right.  \nonumber\\
+\left.  \left.  \left(  2n-5\right)  \kappa^{i}\kappa^{j}g_{ij}^{\left(
0\right)  }\ell^{2}+\left(  2n-3\right)  \ell^{2}\mathcal{R}^{\left(
0\right)  }\right]  +\cdots\right\}  \,.
\end{gather}
Finally, using Eq.(\ref{Tr_Sigma2}), we can rewrite $S_{EE}^{ct}$ as%

\begin{equation}%
\begin{tabular}
[c]{l}%
$S_{EE}^{ct}=-\frac{1}{4G}%
{\displaystyle\int\limits_{\partial\Sigma_{\varepsilon}}}
d^{2n-3}y\frac{\ell\sqrt{\sigma^{\left(  0\right)  }}}{\left(  2n-3\right)
\varepsilon^{\left(  2n-3\right)  /2}}+$\\
$\frac{1}{4G}%
{\displaystyle\int\limits_{\partial\Sigma_{\varepsilon}}}
d^{2n-3}y\frac{\ell^{3}\sqrt{\sigma^{\left(  0\right)  }}}{\varepsilon
^{\left(  2n-5\right)  /2}2\left(  2n-3\right)  ^{2}\left(  2n-5\right)
}\left(  \left(  \frac{2n-5}{2}\right)  \kappa^{i}\kappa^{j}g_{ij}^{\left(
0\right)  }+\frac{\left(  2n-3\right)  ^{2}}{2\left(  2n-4\right)
}\mathcal{R}^{\left(  0\right)  }\right)  +...,$%
\end{tabular}
\ \ \ \ \label{S_EE^counterterm}%
\end{equation}
which explicitly cancells the divergences of $S_{EE}^{RT}$ as presented in
Eq.(\ref{S_EE^RT}), up to next-to-leading order in $\rho=\varepsilon$.

By considering the definition of $S_{EE}^{ren}$, given as the sum of
$S_{EE}^{RT}$ and $S_{EE}^{ct}$, and by using Eq.(\ref{S_EE^RT}) and
Eq.(\ref{S_EE^counterterm}), we finally obtain that $S_{EE}^{ren}=\frac{C_{2}%
}{4G}$, where $C_{2}$ is $O\left(  1\right)  $. Therefore, we have explicitly
verified, up to next-to-leading order, that the definition of renormalized EE
as given in Eq.(\ref{S_EE^ren}) is correct, being finite and equal to the
universal part or the EE. Furthermore, we consider the explicitly verified
cancellation of divergences as evidence of the validity of the generalization
of the Euler theorem to squashed cones of arbitrary even dimensions, which to
the best of our knowledge, has no formal proof yet.

Finally, we note that the EE counterterms can be rewritten in purely intrinsic
form, in terms of the induced metric $\widetilde{\gamma}$ and its curvature
invariants, up to linear order in the curvature. To achieve this, we first
invert the FG expansion of $\sqrt{\widetilde{\gamma}}$, starting from
Eq.(\ref{sqrt_gamma_tilde}), to obtain that%

\begin{equation}
\sqrt{\sigma^{\left(  0\right)  }}=\varepsilon^{\frac{2n-3}{2}}\sqrt
{\widetilde{\gamma}}\left(  1-\frac{1}{2}\varepsilon tr\left[  \sigma^{\left(
2\right)  }\right]  +...\right)  =\varepsilon^{\frac{2n-3}{2}}\sqrt
{\widetilde{\gamma}}-\frac{1}{2}\varepsilon^{\frac{2n-1}{2}}\sqrt
{\widetilde{\gamma}}tr\left[  \sigma^{\left(  2\right)  }\right]  +...
\end{equation}
Then, we replace this into Eq.(\ref{S_EE^counterterm}), and after some
simplifications, which include substituting for $tr\left[  \sigma^{\left(
2\right)  }\right]  $ using Eq.(\ref{Tr_Sigma2}), and relating the Ricci
scalars of $\widetilde{\gamma}_{ab}$ and $\sigma_{ab}^{\left(  0\right)  }$
using Eq.(\ref{Riemann_Gamma}), we have that up to order $\varepsilon
^{-\left(  2n-5\right)  /2}$, $S_{EE}^{ct}$ is given by%

\begin{equation}%
\begin{tabular}
[c]{l}%
$S_{EE}^{ct}=-\frac{1}{4G}%
{\displaystyle\int\limits_{\partial\Sigma_{\varepsilon}}}
d^{2n-3}y\frac{\ell\sqrt{\widetilde{\gamma}}}{\left(  2n-3\right)  }+\frac
{1}{4G}%
{\displaystyle\int\limits_{\partial\Sigma_{\varepsilon}}}
d^{2n-3}y\frac{\ell^{3}\sqrt{\widetilde{\gamma}}}{2\left(  2n-3\right)
\left(  2n-4\right)  \left(  2n-5\right)  }\mathcal{R}\left[  \widetilde
{\gamma}_{ab}\right]  +...,$%
\end{tabular}
\ \ \ \ \ \ \ \ \label{S_EE^ren_intrinsic}%
\end{equation}
which is written in terms of purely intrinsic quantities. This intrinsic form
of the counterterms allows to make contact with the EE renormalization scheme
presented in Ref.\cite{Marika}. In appendix \ref{AppB}, we show how it can be
derived starting directly from $S_{EE}^{ct}$ written in terms of the Chern
form, as given in Eq.(\ref{S_EE^ct}).

\subsection{Topological interpretation of renormalized EE for the AdS$_{2n}%
$/CFT$_{2n-1}$ case\label{TopolInterp}}

We now reinterpret $S_{EE}^{ren}$ as given in Eq.(\ref{S_EE^ren}), in terms of
the topological and geometric properties of the minimal surface $\Sigma$ as an
AAdS submanifold. In particular, we rewrite $S_{EE}^{ren}$ in terms of the
Euler characteristic of $\Sigma$ and its AdS curvature \cite{OleaF}.

By considering the Euler theorem, as presented in Eq.(\ref{Usual_Euler}), we
have that the Chern form which gives the EE counterterm can be rewritten as%

\begin{equation}%
{\displaystyle\int\limits_{\partial\Sigma}}
B_{2n-3}=%
{\displaystyle\int\limits_{\Sigma}}
d^{2n-2}y\sqrt{\gamma}\varepsilon_{2n-2}-\left(  4\pi\right)  ^{\left(
n-1\right)  }\left(  n-1\right)  !\chi\left[  \Sigma\right]  ,
\end{equation}
where $\varepsilon_{2n-2}$ is the Euler density of $\Sigma$ and $\chi\left[
\Sigma\right]  $ is its Euler characteristic. Then, $S_{EE}^{ren}$ can be
expressed as%

\begin{gather}
S_{EE}^{ren}=\frac{1}{4G}\left(  {\displaystyle\int\limits_{\Sigma}}%
d^{2n-2}y\sqrt{\gamma}+\frac{\left(  -1\right)  ^{n}\ell^{2\left(  n-1\right)
}}{\left(  2n-2\right)  !}{\displaystyle\int\limits_{\Sigma}}d^{2n-2}%
y\sqrt{\gamma}\varepsilon_{2n-2}\right)  +\nonumber\\
+\frac{\left(  -1\right)  ^{n+1}\left(  4\pi\right)  ^{\left(  n-1\right)
}\left(  n-1\right)  !\ell^{2\left(  n-1\right)  }}{4G\left(  2n-2\right)
!}\chi\left[  \Sigma\right]  \,.
\end{gather}

We now define the topological constant $C_{\chi}$ as%

\begin{equation}
C_{\chi}=\frac{\left(  -1\right)  ^{n+1}\left(  4\pi\right)  ^{\left(
n-1\right)  }\left(  n-1\right)  !\ell^{2\left(  n-1\right)  }}{4G\left(
2n-2\right)  !}\chi\left[  \Sigma\right]  ,\label{Topol_const}%
\end{equation}
and using the definition of $\varepsilon_{2n-2}$ given in Eq.(\ref{Euler_2n}),
we can write $S_{EE}^{ren}$ as%

\begin{equation}
S_{EE}^{ren}=\frac{1}{4G}{\displaystyle\int\limits_{\Sigma}}d^{2n-2}%
y\sqrt{\gamma}\left(  1+\frac{\left(  -1\right)  ^{n}\ell^{2\left(
n-1\right)  }}{\left(  2n-2\right)  !2^{\left(  n-1\right)  }}\delta_{\left[
a_{1}...a_{2n-2}\right]  }^{\left[  b_{1}...b_{2n-2}\right]  }\mathcal{R}%
_{b_{1}b_{2}}^{a_{1}a_{2}}...\mathcal{R}_{b_{2n-3}b_{2n-2}}^{a_{2n-3}a_{2n-2}%
}\right)  +C_{\chi}\,,\label{S_partial_step1}%
\end{equation}
where $\mathcal{R}_{b_{1}b_{2}}^{a_{1}a_{2}}$ is the Riemann tensor of the
induced metric $\gamma_{ab}$ on $\Sigma$. Finally, we can simplify
Eq.(\ref{S_partial_step1}) by considering the properties of the antisymmetric
Kronecker delta. In particular, we have that%

\begin{gather}
S_{EE}^{ren}=\frac{\ell^{2\left(  n-1\right)  }}{4G2^{\left(  n-1\right)
}\left(  2n-2\right)  !} {\displaystyle\int\limits_{\Sigma}} d^{2n-2}%
y\sqrt{\gamma}\delta_{\left[  a_{1}...a_{2n-2}\right]  }^{\left[
b_{1}...b_{2n-2}\right]  } \left[  \left(  -1\right)  ^{n}\mathcal{R}%
_{b_{1}b_{2}}^{a_{1}a_{2}}...\mathcal{R} _{b_{2n-3}b_{2n-2}}^{a_{2n-3}%
a_{2n-2}} + \right. \nonumber\\
+ \left.  \frac{1}{\ell^{2\left(  n-1\right)  }} \delta_{\left[  b_{1}%
b_{2}\right]  }^{\left[  a_{1}a_{2}\right]  }...\delta_{\left[  b_{2n-3}%
b_{2n-2}\right]  }^{\left[  a_{2n-3}a_{2n-2}\right]  }\right]  + C_{\chi} \,.
\label{S_partial_step2}%
\end{gather}

Now, we express $S_{EE}^{ren}$ in terms of the AdS curvature $\mathcal{F}%
_{AdS}$, which for a general AAdS manifold is defined as%

\begin{equation}
\left(  \mathcal{F}_{AdS}\right)  _{\nu_{1}\nu_{2}}^{\mu_{1}\mu_{2}}%
=R_{\nu_{1}\nu_{2}}^{\mu_{1}\mu_{2}}+\frac{1}{\ell^{2}}\delta_{\left[  \nu
_{1}\nu_{2}\right]  }^{\left[  \mu_{1}\mu_{2}\right]  }, \label{AdS_curvature}%
\end{equation}
where $R_{\nu_{1}\nu_{2}}^{\mu_{1}\mu_{2}}$ is the Riemann tensor of the
manifold. Then, for the $\Sigma$ manifold, the product of the Riemann tensors
can be reexpressed in terms of the AdS curvature as%

\begin{equation}%
\begin{tabular}
[c]{l}%
$\mathcal{R}_{b_{1}b_{2}}^{a_{1}a_{2}}...\mathcal{R}_{b_{2n-3}b_{2n-2}%
}^{a_{2n-3}a_{2n-2}}=\left(  \left(  \mathcal{F}_{\left.  AdS\right\vert
_{\Sigma}}\right)  _{b_{1}b_{2}}^{a_{1}a_{2}}-\frac{1}{\ell^{2}}%
\delta_{\left[  b_{1}b_{2}\right]  }^{\left[  a_{1}a_{2}\right]  }\right)
\cdots\left(  \left(  \mathcal{F}_{\left.  AdS\right\vert _{\Sigma}}\right)
_{b_{2n-3}b_{2n-2}}^{a_{2n-3}a_{2n-2}}-\frac{1}{\ell^{2}}\delta_{\left[
b_{2n-3}b_{2n-2}\right]  }^{\left[  a_{2n-3}a_{2n-2}\right]  }\right)  $\\
$=%
{\displaystyle\sum\limits_{k=0}^{n-1}}
\binom{n-1}{k}\left(  -\frac{1}{\ell^{2}}\right)  ^{\left(  n-1-k\right)
}\left[  \left(  \mathcal{F}_{\left.  AdS\right\vert _{\Sigma}}\right)
_{b_{1}b_{2}}^{a_{1}a_{2}}\cdots\left(  \mathcal{F}_{\left.  AdS\right\vert
_{\Sigma}}\right)  _{b_{2k-1}b_{2k}}^{a_{2k-1}a_{2k}}\right]  \left[
\delta_{\left[  b_{2k+1}b_{2k+2}\right]  }^{\left[  a_{2k+1}a_{2k+2}\right]
}\cdots\delta_{\left[  b_{2n-3}b_{2n-2}\right]  }^{\left[  a_{2n-3}%
a_{2n-2}\right]  }\right]  ,$%
\end{tabular}
\end{equation}
where we have disregarded the order of the indices due to the presence of the
overall Kronecker delta. Now, noting that the $k=0$ term cancells the deltas
appearing in Eq.(\ref{S_partial_step2}), we can write%

\begin{gather}
S_{EE}^{ren}={\displaystyle\int\limits_{\Sigma}}\frac{d^{2n-2}y\sqrt{\gamma
}\ell^{2\left(  n-1\right)  }}{4G2^{\left(  n-1\right)  }\left(  2n-2\right)
!} \delta_{\left[  a_{1}...a_{2n-2}\right]  }^{\left[  b_{1}...b_{2n-2}%
\right]  }{\displaystyle\sum\limits_{k=1}^{n-1}} \frac{\left(  -1\right)
^{1+k}}{\ell^{2\left(  n-1-k\right)  }}\binom{n-1} {k} \left(  \mathcal{F}%
_{\left.  AdS\right\vert _{\Sigma}}\right)  _{b_{1}b_{2}}^{a_{1}a_{2}}
\ldots\left(  \mathcal{F}_{\left.  AdS\right\vert _{\Sigma}}\right)
_{b_{2k-1}b_{2k}}^{a_{2k-1}a_{2k}} \times\nonumber\\
\times\delta_{\left[  b_{2k+1}b_{2k+2}\right]  }^{\left[  a_{2k+1}%
a_{2k+2}\right]  }\ldots\delta_{\left[  b_{2n-3}b_{2n-2}\right]  }^{\left[
a_{2n-3}a_{2n-2}\right]  } + C_{\chi} \,.
\end{gather}

Finally, using the properties of the antisymmetric delta, we obtain that%

\begin{gather}
S_{EE}^{ren}={\displaystyle\int\limits_{\Sigma}}\frac{d^{2n-2}y\sqrt{\gamma
}\ell^{2\left(  n-1\right)  }}{4G2^{\left(  n-1\right)  }\left(  2n-2\right)
!}{\displaystyle\sum\limits_{k=1}^{n-1}}\frac{\left(  -1\right)  ^{\left(
1+k\right)  }\left[  2\left(  n-1-k\right)  \right]  !2^{\left(  n-1-k\right)
}}{\ell^{2\left(  n-1-k\right)  }}\binom{n-1}{k}\times\nonumber\\
\times\delta_{\left[  a_{1}...a_{2k}\right]  }^{\left[  b_{1}...b_{2k}\right]
}\left(  \mathcal{F}_{\left.  AdS\right\vert _{\Sigma}}\right)  _{b_{1}b_{2}%
}^{a_{1}a_{2}}\ldots\left(  \mathcal{F}_{\left.  AdS\right\vert _{\Sigma}%
}\right)  _{b_{2k-1}b_{2k}}^{a_{2k-1}a_{2k}}+C_{\chi}\,,\label{S_EE^Topol}%
\end{gather}
where $S_{EE}^{ren}$ has been expressed in terms of contractions of the AdS
curvature of $\Sigma$ ($\mathcal{F}_{\left.  AdS\right\vert _{\Sigma}}$) and
its Euler characteristic, as considered in the definition of $C_{\chi}$ given
in Eq.(\ref{Topol_const}).

We note that Eq.(\ref{S_EE^Topol}) gives an interpretation of renormalized EE
in terms of the topological and geometric properties of $\Sigma$ as an AAdS
Riemannian manifold, which generalizes the result obtained in Ref.\cite{AAO1}
for the case of AdS$_{4}$/CFT$_{3}$.

\section{Explicit example: Ball-shaped entangling region in CFT$_{2n-1}$, with
a global AdS$_{2n}$ bulk\label{Explicit}}

In order to exhibit the advantages of the topological renormalization scheme,
we now rederive the renormalized EE of a ball-shaped entangling region in the
ground state of an holographic $\left(  2n-1\right)  -$dimensional CFT, having
global AdS$_{2n}$ as its gravity dual. This computation was firstly done by
Kawano, Nakaguchi and Nishioka in Ref.\cite{NishiokaPaper}, by evaluating the
RT area functional. Instead, we make use of the topological form for
$S_{EE}^{ren}$ given in Eq.(\ref{S_EE^Topol}).

We consider that the entangling region is delimited by a spherical entangling
surface, such that $\partial A=S^{2n-3}$. Also, the metric $G_{\mu\nu}$ of
global AdS$_{2n}$ can be written as%

\begin{equation}%
\begin{tabular}
[c]{l}%
$ds_{G}^{2}=\frac{\ell^{2}d\rho^{2}}{4\rho^{2}}+\frac{1}{\rho}\left(
-dt^{2}+dr^{2}+r^{2}d\Omega_{2n-3}^{2}\right)  =G_{\mu\nu}dx^{\mu}dx^{\nu},$\\
$d\Omega_{2n-3}^{2}=d\theta_{1}^{2}+\sin^{2}\theta_{1}d\theta_{2}^{2}%
+...+\sin^{2}\theta_{1}\cdots\sin^{2}\theta_{2n-4}d\theta_{2n-3}^{2},$%
\end{tabular}
\ \label{Metric_G}%
\end{equation}
where the boundary metric has been expressed in spherical coordinates. Now, as
shown in appendix \ref{Verification}, the minimal surface $\Sigma$
corresponding to the entangling surface $\partial A=S^{2n-3}$ can be
parametrized as%

\begin{equation}
\Sigma:\left\{  t=const~;~r^{2}+\ell^{2}\rho=R^{2}\right\}  ,
\label{Minimal_Surface_Param}%
\end{equation}
where $R$ is the radius of the sphere. We proceed to compute the induced
metric $\gamma_{ab}$ on $\Sigma$, defined in Eq.(\ref{InducedMetricSigma}),
considering that its worldvoume coordinates are given by $y^{a}=\left\{
\rho,\theta_{1},...,\theta_{2n-3}\right\}  $, whereas the bulk coordinates are
$x^{\mu}=\left\{  \rho,t,r,\theta_{1},...,\theta_{2n-3}\right\}  $. In
particular, we obtain that%

\begin{equation}
ds_{\gamma}^{2}=\frac{\ell^{2}}{4\rho^{2}}\left(  1+\frac{\ell^{2}\rho
}{\left(  R^{2}-\ell^{2}\rho\right)  }\right)  d\rho^{2}+\frac{\left(
R^{2}-\ell^{2}\rho\right)  }{\rho}d\Omega_{2n-3}^{2}=\gamma_{ab}dy^{a}dy^{b}.
\label{Induced_Metric_explicit}%
\end{equation}

Considering the induced metric, we now compute $S_{EE}^{ren}$ using the
topological procedure. Given the induced metric of
Eq.(\ref{Induced_Metric_explicit}), we compute the AdS curvature on $\Sigma$,
according to Eq.(\ref{AdS_curvature}), and we find that it vanishes
identically (i.e., $\left(  \mathcal{F}_{\left.  AdS\right\vert _{\Sigma}%
}\right)  _{b_{1}b_{2}}^{a_{1}a_{2}}=0$). Also, we consider that $\Sigma$ is
topologically equivalent to a $\left(  2n-2\right)  -$ball, whose Euler
characteristic is $\chi\left(  \Sigma\right)  =1$. Therefore, using
$S_{EE}^{ren}$ as given in Eq.(\ref{S_EE^Topol}), we have that
\begin{equation}
S_{EE}^{ren}=\frac{\left(  -1\right)  ^{n+1}\left(  4\pi\right)  ^{\left(
n-1\right)  }\left(  n-1\right)  !\ell^{2\left(  n-1\right)  }}{4G\left(
2n-2\right)  !},\label{S_EE^ren_answer}%
\end{equation}
which agrees with the standard result as given in
Refs.\cite{NishiokaPaper,NishiokaReview}. We also conclude from this analysis
that for spherical entangling surfaces, the minimal surface $\Sigma$ is a
constant curvature surface, which has $\mathcal{R}_{b_{1}b_{2}}^{a_{1}a_{2}%
}=-\frac{1}{\ell^{2}}\delta_{\left[  b_{1}b_{2}\right]  }^{\left[  a_{1}%
a_{2}\right]  }$.

We note that the computation of the renormalized EE using the topological
approach of Eq.(\ref{S_EE^Topol}) is performed directly to all orders in
$\rho=\varepsilon$. We also mention that, as further explained in section
\ref{Outlook}, the computation of $S_{EE}^{ren}$ for the case with $\partial
A=S^{2n-3}$ is important because it is related to the $a-$ charge
\cite{S-c-Myers,NishiokaPaper}, such that $S_{EE}^{ren}=\left(  -1\right)
^{n-1}2\pi a_{2n-1}$. The $a-$charge counts the number of degrees of freedom
of the CFT and is conjectured to decrease along RG flows between conformal
fixed points. Therefore, it can be thought of as a generalization of
Zamolodchikov's c-theorem \cite{cTheo}.

Using our result for $S_{EE}^{ren}$, we therefore find that%

\begin{equation}
a_{2n-1}=\frac{\left(  4\pi\right)  ^{\left(  n-2\right)  }\left(  n-1\right)
!\ell^{2\left(  n-1\right)  }}{2G\left(  2n-2\right)  !},\label{a-charge}%
\end{equation}
in agreement with the known result as given in
Refs.\cite{NishiokaPaper,NishiokaReview}. As written, the $a-$charge is
expressed in terms of the bulk gravity quantities (like the AdS radius $\ell$
and Newton's constant $G$), but it can be rewritten entirely in terms of the
CFT quantities by using the standard dictionary.

\section{Outlook\label{Outlook}}

We have successfully extended the topological scheme developed in
Ref.\cite{AAO1} for computing the renormalized EE to holographic CFTs of
arbitrary odd dimensions. This procedure considers adding the Chern form to
the usual RT area functional, such that the renormalized EE is written as
shown in Eq.(\ref{S_EE^ren}). Alternatively, $S_{EE}^{ren}$ can be written in
terms of the Euler characteristic of the minimal surface $\Sigma$ and its AdS
curvature, as shown in Eq.(\ref{S_EE^Topol}). The latter form greatly
simplifies the computation of $S_{EE}^{ren}$ and is of interest because it
exhibits the relation between the EE and the topological and geometric
properties of $\Sigma$. We also make contact with the renormalization
procedure developed in Ref.\cite{Marika}, by writing the EE counterterm in
terms of the intrinsic quantities on $\partial\Sigma$, as shown in
Eq.(\ref{S_EE^ren_intrinsic}).

The renormalized EE is of interest for the study of holographic
renormalization group (RG) flows. As firstly mentioned in Ref.\cite{Marika}
and also discussed in Ref.\cite{NishiokaReview}, $S_{EE}^{ren}$ for a
ball-shaped entangling region in the ground state of a CFT is related to its
$a-$charge \cite{S-c-Myers}, which encodes information about the number of
degrees of freedom of the theory and is conjectured to decrease along RG flows
between conformal fixed points. It therefore constitutes a generalization of
Zamolodchikov's c-theorem \cite{cTheo}. In particular, for a $\left(
2n-1\right)  -$dimensional CFT, $S_{EE}^{ren}=\left(  -1\right)  ^{n-1}2\pi
a_{2n-1}$ where $\left(  a_{2n-1}\right)  _{UV}\geq\left(  a_{2n-1}\right)
_{IR}$ between any two fixed points. Using our topological renormalization
approach (see Eq.(\ref{S_EE^Topol})), the computation of $S_{EE}^{ren}$ for
spherical entangling surfaces becomes nearly trivial, and as discussed in
section \ref{Explicit}, we recover the known result given in
Refs.\cite{NishiokaPaper,NishiokaReview}.

As future work, we will also study how to extend the scheme to AAdS manifolds
of arbitrary odd dimensions by using the renormalized Euclidean gravitational
action discussed in Ref.\cite{K2Odd}. We expect this analysis to be useful for
the study of the conformal anomaly of the corresponding holographic CFTs. We
also intend to apply our topological renormalization scheme to obtain the
renormalized EE for higher-curvature theories of gravity, specially those of
the Lovelock class \cite{Lovelock1,Lovelock2}, and to renormalize other
information-theoretic measures of holographic CFTs, like the Entanglement
Renyi Entropies (EREs)
\cite{RenyiXiDong,RenyiHeadrick,RenyiFreeEnergy,MyersRenyi} and the complexity
\cite{AliComplexity,Complexity2,Complexity3,IgnacioReyes}.

\begin{acknowledgments}
The authors thank C. Arias and Y. Novoa for interesting discussions. We also
thank S. N. Solodukhin for relevant correspondence. G.A. is a Universidad
Andres Bello (UNAB) Ph.D. Scholarship holder, and his work is supported by
Direcci\'{o}n General de Investigaci\'{o}n (DGI-UNAB). This work is funded in
part by FONDECYT Grants No. 1170765 and No. 3180620, UNAB Grant DI-1336-16/R
and CONICYT Grant DPI 20140115.
\end{acknowledgments}

\appendix

\section{Explicit expansion of the Chern form\label{AppA}}

We proceed to simplify and expand the $B_{2n-3}$ Chern form, which appears in
the expression for $S_{EE}^{ct}$ as given in Eq.(\ref{S_EE^ct}). In the
following expressions we introduce a short-hand notation, such that the
antisymmetric Kronecker deltas are indicated only by the number of indices,
i.e., $\delta_{\lbrack b_{1}...b_{2n-3}]}^{[a_{1}...a_{2n-3}]}$ is denoted by
$\delta_{\lbrack2n-3]}^{[2n-3]}$. Also, as the final expression is fully
contracted, the indices of all tensors are omitted. For example, $k_{b_{1}%
}^{a_{1}}k_{b_{2}}^{a_{2}}$ is denoted by $k^{2}$, and $\mathcal{R}%
_{b_{1}b_{2}}^{a_{1}a_{2}}$, which is the Riemann tensor of the metric
$\widetilde{\gamma}$, is denoted as $\mathcal{R}ie$. To avoid ambiguity,
traces of tensors are explicitly indicated.

In the new notation, the definition of the Chern form at $\partial\Sigma$, is
given by%

\begin{equation}
B_{2n-3}=-2\left(  n-1\right)  {\displaystyle\int\limits_{0}^{1}}%
dtd^{2n-3}y\sqrt{\widetilde{\gamma}}\delta_{\left[  2n-3\right]  }^{\left[
2n-3\right]  }k\left(  \frac{1}{2}\mathcal{R}ie-t^{2}k^{2}\right)  ^{n-2},
\label{B2n-3Appendix}%
\end{equation}
and the FG expansion of the different quantities are given in section
\ref{Finiteness}. We note that, in every FG expansion, the $"..."$ represents
$O\left(  \rho^{2}\right)  $ terms.

First, we expand the $k^{2}$ terms, were we have that%

\begin{equation}%
\begin{tabular}
[c]{l}%
$k^{2}=\frac{1}{\ell^{2}}\left(  \delta^{2}-2\rho\left[  \delta\sigma^{\left(
2\right)  }+\frac{\ell^{2}\kappa^{i}\kappa^{j}g_{ij}^{\left(  0\right)  }%
}{2\left(  2n-3\right)  ^{2}}\delta^{2}\right]  +...\right)  .$%
\end{tabular}
\end{equation}
Then, we consider that the term in parenthesis can be expanded as%

\begin{equation}
\left(  \frac{1}{2}\mathcal{R}ie-t^{2}k^{2}\right)  =-\frac{t^{2}}{\ell^{2}%
}\delta^{2}+\rho\left[  \frac{1}{2}\left(  \mathcal{R}ie^{\left(  0\right)
}\right)  +\frac{2}{l^{2}}t^{2}\left(  \delta\sigma^{\left(  2\right)  }%
+\frac{\ell^{2}\kappa^{i}\kappa^{j}g_{ij}^{\left(  0\right)  }}{2\left(
2n-3\right)  ^{2}}\delta^{2}\right)  \right]  +\ldots\,.\nonumber
\end{equation}
Note that $\mathcal{R}ie^{\left(  0\right)  }$ is the Riemann tensor of the
metric $\sigma^{\left(  0\right)  }$. The product of such terms, to linear
order in $\rho$, becomes%

\begin{gather}
\left(  \frac{1}{2}\mathcal{R}ie-t^{2}k^{2}\right)  ^{n-2}=\left(  -1\right)
^{n-2}\frac{t^{2n-4}}{\ell^{2n-4}}\delta^{2n-4}+\nonumber\\
\rho\left(  -1\right)  ^{n-3}\left(  n-2\right)  \frac{t^{2n-6}}{\ell^{2n-6}%
}\delta^{2n-6}\left[  \frac{1}{2}\left(  \mathcal{R}ie^{\left(  0\right)
}\right)  +\frac{2}{\ell^{2}}t^{2}\left(  \delta\sigma^{\left(  2\right)
}+\frac{\ell^{2}\kappa^{i}\kappa^{j}g_{ij}^{\left(  0\right)  }}{2\left(
2n-3\right)  ^{2}}\delta^{2}\right)  \right]  +\ldots\,.
\end{gather}
Therefore, $\sqrt{\widetilde{\gamma}}k\left(  \frac{1}{2}\mathcal{R}%
ie-t^{2}k^{2}\right)  ^{n-2}$ can be expanded as%

\begin{gather}
\sqrt{\widetilde{\gamma}}k\left(  \frac{1}{2}\mathcal{R}ie-t^{2}k^{2}\right)
^{n-2}=\frac{\left(  -1\right)  ^{\left(  n-2\right)  }\sqrt{\sigma
_{\alpha\beta}^{\left(  0\right)  }}}{\ell^{2n-3}\rho^{\left(  2n-3\right)
/2}}t^{2n-4}\delta^{2n-3}\nonumber\\
-\frac{\left(  -1\right)  ^{\left(  n-2\right)  }\sqrt{\sigma_{\alpha\beta
}^{\left(  0\right)  }}}{\ell^{2n-3}\rho^{\left(  2n-3\right)  /2}}%
t^{2n-4}\rho\left[  -\frac{tr[\sigma^{\left(  2\right)  }]}{2}\delta
^{2n-3}+\left(  2n-3\right)  \delta^{2n-4}\left(  \sigma^{\left(  2\right)
}\right)  \right.  \nonumber\\
+\left.  \frac{\left(  n-2\right)  \ell^{2}}{2t^{2}}\delta^{2n-5}\left(
\mathcal{R}ie^{\left(  0\right)  }\right)  +\frac{\kappa^{i}\kappa^{j}%
g_{ij}^{\left(  0\right)  }\ell^{2}}{2\left(  2n-3\right)  }\delta
^{2n-3}\right]  +...~.
\end{gather}
Thus, the t integral of the previous term has a FG expansion given by%

\begin{gather}
{\displaystyle\int\limits_{0}^{1}}dt\sqrt{\widetilde{\gamma}}k\left(  \frac
{1}{2}\mathcal{R}ie-t^{2}k^{2}\right)  ^{n-2}=\frac{\left(  -1\right)
^{n-2}\sqrt{\sigma_{\alpha\beta}^{\left(  0\right)  }}}{\ell^{2n-3}%
\rho^{\left(  2n-3\right)  /2}}\frac{1}{2n-3}\delta^{2n-3}\nonumber\\
-\frac{\left(  -1\right)  ^{n-2}\sqrt{\sigma_{\alpha\beta}^{\left(  0\right)
}}}{\ell^{2n-3}\rho^{\left(  2n-3\right)  /2}}\frac{1}{2n-3}\rho\left[
-\frac{tr[\sigma^{\left(  2\right)  }]}{2}\delta^{2n-3}+\left(  2n-3\right)
\delta^{2n-4}\sigma^{\left(  2\right)  }\right. \nonumber\\
+\left.  \frac{\left(  n-2\right)  \left(  2n-3\right)  \ell^{2}}{2\left(
2n-5\right)  }\delta^{2n-5}\left(  \mathcal{R}ie^{\left(  0\right)  }\right)
+\frac{\kappa^{i}\kappa^{j}g_{ij}^{\left(  0\right)  }\ell^{2}}{2\left(
2n-3\right)  }\delta^{2n-3}\right]  +\ldots~.
\end{gather}

Now, using that%

\begin{equation}%
\begin{tabular}
[c]{l}%
$\delta_{\lbrack j_{1}...j_{m}]}^{[i_{1}...i_{m}]}\delta_{i_{1}}^{j_{1}%
}...\delta_{i_{k}}^{j_{k}}=\frac{\left(  N-m+k\right)  !}{\left(  N-m\right)
!}\delta_{\left[  j_{k+1}...j_{m}\right]  }^{\left[  i_{k+1}...i_{m}\right]
},$%
\end{tabular}
\end{equation}
we can finally obtain the contractions of the integrand, such that%

\begin{gather}
{\displaystyle\int\limits_{0}^{1}}dt\sqrt{\widetilde{\gamma}}\delta_{\left[
2n-3\right]  }^{\left[  2n-3\right]  }k\left(  \frac{1}{2}\mathcal{R}%
ie-t^{2}k^{2}\right)  ^{n-2}=\frac{\left(  -1\right)  ^{n-2}\left(
2n-3\right)  !\sqrt{\sigma_{\alpha\beta}^{\left(  0\right)  }}}{\ell
^{2n-3}\left(  2n-3\right)  \rho^{\left(  2n-3\right)  /2}}\nonumber\\
-\frac{\left(  -1\right)  ^{n-2}\left(  2n-3\right)  !\sqrt{\sigma
_{\alpha\beta}^{\left(  0\right)  }}}{\ell^{2n-3}\left(  2n-3\right)
\rho^{\left(  2n-3\right)  /2}}\frac{\rho}{2\left(  2n-3\right)  \left(
2n-5\right)  }\left[  \left(  2n-3\right)  \left(  2n-5\right)  tr[\sigma
^{\left(  2\right)  }]\right. \nonumber\\
+\left.  \left(  2n-5\right)  \kappa^{i}\kappa^{j}g_{ij}^{\left(  0\right)
}\ell^{2}+\left(  2n-3\right)  \ell^{2}\mathcal{R}^{\left(  0\right)
}\right]  +\ldots~,
\end{gather}
where $\mathcal{R}^{\left(  0\right)  }$ is the Ricci scalar of the
$\sigma^{\left(  0\right)  }$ metric.

Finally, replacing the previous expression into Eq.(\ref{B2n-3Appendix}), we
obtain that $B_{2n-3}$ is given by%

\begin{gather}
B_{2n-3}=\frac{2\left(  n-1\right)  \left(  -1\right)  ^{n-1}\left(
2n-2\right)  !}{\ell^{2n-3}}\frac{d^{2n-3}y\sqrt{\sigma_{\alpha\beta}^{\left(
0\right)  }}}{\varepsilon^{\left(  2n-3\right)  /2}}\left\{  1-\frac
{\varepsilon}{2\left(  2n-3\right)  \left(  2n-5\right)  }\left[  \left(
2n-3\right)  \left(  2n-5\right)  tr[\sigma^{\left(  2\right)  }]\right.
\right.  \nonumber\\
+\left.  \left.  \left(  2n-5\right)  \kappa^{i}\kappa^{j}g_{ij}^{\left(
0\right)  }\ell^{2}+\left(  2n-3\right)  \ell^{2}\mathcal{R}^{\left(
0\right)  }\right]  +\ldots\right\}
\end{gather}
which corresponds to Eq.(\ref{B2n-3MainText}) of section \ref{Finiteness} of
the main text.

\section{Intrinsic counterterms directly from the Chern form\label{AppB}}

In this appendix we emphasize that the intrinsic counterterms for the EE, as
presented in Eq.(\ref{S_EE^ren_intrinsic}), can be directly obtained starting
from the Chern form at $\partial\Sigma$. In particular, considering
Eq.(\ref{S_EE^ct}), we have that%

\begin{equation}%
\begin{tabular}
[c]{l}%
$S_{EE}^{ct}=\frac{\left(  -1\right)  ^{n}\ell^{2\left(  n-1\right)  }%
}{4G\left[  2\left(  n-1\right)  \right]  !}%
{\displaystyle\int\limits_{\partial\Sigma}}
B_{2n-3},$%
\end{tabular}
\ \ \ \label{Counter-appendix}%
\end{equation}
where $B_{2n-3}$ is given in Eq.(\ref{B2n-3Appendix}). We use the same
shorthand notation introduced in appendix \ref{AppA}.

Now, the Riemann tensor of the $\widetilde{\gamma}_{ab}$ can be written in
terms of the Weyl tensor and the Schouten tensor, such that%

\begin{equation}
\mathcal{R}_{b_{1}b_{2}}^{a_{1}a_{2}}=\mathcal{W}_{b_{1}b_{2}}^{a_{1}a_{2}%
}+\delta_{\lbrack b_{1}}^{[a_{1}}\mathcal{S}_{b_{2}]}^{a_{2}]},
\end{equation}
where the antisymmetrization does not have a $\frac{1}{4}$ factor. In turn,
the Schouten and the Weyl tensors of the metric $\widetilde{\gamma}_{ab}$ at
$\partial\Sigma$ can be expressed in terms of those of the $\sigma
_{ab}^{\left(  0\right)  }$ metric at $\partial A$, to first order in $\rho$, as%

\begin{equation}%
\begin{tabular}
[c]{l}%
$\mathcal{S}_{b}^{a}=\rho\left(  S^{\left(  0\right)  }\right)  _{b}%
^{a}+...~,$\\
$\mathcal{W}_{b_{1}b_{2}}^{a_{1}a_{2}}=\rho\left(  W^{\left(  0\right)
}\right)  _{b_{1}b_{2}}^{a_{1}a_{2}}+...~.$%
\end{tabular}
\end{equation}
Then, considering the FG expansion of the extrinsic curvature $k_{b}^{a}$ of
$\partial\Sigma$ as given in Eq.(\ref{k-expansion}) and using that $d=2n-1$,
we have that%

\begin{equation}%
\begin{tabular}
[c]{l}%
$k_{b}^{a}=\left(  k^{\left(  0\right)  }\right)  _{b}^{a}+\rho\left(
k^{\left(  2\right)  }\right)  _{b}^{a}+...,$\\
$~\left(  k^{\left(  0\right)  }\right)  _{b}^{a}=\frac{1}{\ell}\delta_{b}%
^{a}~,$\\
$\left(  k^{\left(  2\right)  }\right)  _{b}^{a}=-\frac{1}{\ell}\left[
\left(  \sigma^{\left(  2\right)  }\right)  _{b}^{a}+\frac{\ell^{2}\kappa
^{i}\kappa^{j}g_{ij}^{\left(  0\right)  }}{2\left(  2n-3\right)  ^{2}}%
\delta_{b}^{a}\right]  .$%
\end{tabular}
\end{equation}
Finally, using that%
\begin{equation}
tr\left[  \sigma^{\left(  2\right)  }\right]  =-\ell^{2}S\left[
\sigma^{\left(  0\right)  }\right]  -\frac{\ell^{2}}{2\left(  2n-3\right)
}\kappa^{i}\kappa^{j}g_{ij}^{\left(  0\right)  },
\end{equation}
as given in Eq.(\ref{Trace-of-sigma2}), we have that%

\begin{equation}
tr[k^{\left(  2\right)  }]=\ell S\left[  \sigma^{\left(  0\right)  }\right]
=\ell S^{\left(  0\right)  }.
\end{equation}

We proceed to simplify the expression for $B_{2n-3}$. Using the shorthand
notation, we have that%

\begin{equation}%
\begin{tabular}
[c]{l}%
$k^{2}=\left(  k^{\left(  0\right)  }\right)  +2\rho k^{\left(  0\right)
}k^{\left(  2\right)  }+...,$\\
$\frac{1}{2}\delta_{\left[  2\right]  }^{\left[  2\right]  }\mathcal{R}%
ie=\frac{1}{2}\rho\left(  W^{\left(  0\right)  }+4\delta S^{\left(  0\right)
}\right)  +...,$%
\end{tabular}
\end{equation}
and also%

\begin{equation}
\left(  \frac{1}{2}\mathcal{R}ie-t^{2}k^{2}\right)  =-t^{2}\left(  k^{\left(
0\right)  }\right)  ^{2}+\rho\left[  \frac{1}{2}W^{\left(  0\right)  }+2\delta
S^{\left(  0\right)  }-2t^{2}k^{\left(  2\right)  }k^{\left(  0\right)
}\right]  +...\,.\nonumber
\end{equation}
Now, the product of the previous terms can be expanded as%

\begin{gather}
\left(  \frac{1}{2}\mathcal{R}ie-t^{2}k^{2}\right)  ^{n-2}=\left(
-t^{2}\right)  ^{\left(  n-2\right)  }\left(  k^{\left(  0\right)  }\right)
^{2n-4}\nonumber\\
+\rho\left(  n-2\right)  \left(  -t^{2}\right)  ^{\left(  n-3\right)  }\left(
k^{\left(  0\right)  }\right)  ^{2n-6}\left[  \frac{1}{2}W^{\left(  0\right)
}+2S^{\left(  0\right)  }\delta-2t^{2}k^{\left(  2\right)  }k^{\left(
0\right)  }\right]  +...~,
\end{gather}
and thus%

\begin{gather}
k\left(  \frac{1}{2}\mathcal{R}ie-t^{2}k^{2}\right)  ^{n-2}=\left(  -1\right)
^{n-2}t^{2n-4}\left(  k^{\left(  0\right)  }\right)  ^{2n-3}\nonumber\\
+\left(  -1\right)  ^{n-2}\rho\left[  -\left(  2n-4\right)  t^{2n-6}\left(
k^{\left(  0\right)  }\right)  ^{2n-5}\delta S^{\left(  0\right)  }+\left(
2n-3\right)  t^{2n-4}\left(  k^{\left(  0\right)  }\right)  ^{2n-4}k^{\left(
2\right)  }\right]  +...,
\end{gather}
where we used the fact that the contractions of the Weyl tensor, after
considering the overall Kronecker delta, vanish identically due to its
tracelessnes. Now, using that $\left(  k^{\left(  0\right)  }\right)  _{b}%
^{a}=\frac{1}{\ell}\delta_{b}^{a}$, $tr[k^{\left(  2\right)  }]=\ell tr\left[
S^{\left(  0\right)  }\right]  $ and%

\begin{equation}%
\begin{tabular}
[c]{l}%
$\delta_{\lbrack j_{1}...j_{m}]}^{[i_{1}...i_{m}]}\delta_{i_{1}}^{j_{1}%
}...\delta_{i_{k}}^{j_{k}}=\frac{\left(  N-m+k\right)  !}{\left(  N-m\right)
!}\delta_{\left[  j_{k+1}...j_{m}\right]  }^{\left[  i_{k+1}...i_{m}\right]
},$%
\end{tabular}
\end{equation}
we can compute the contractions with the overall Kronecker delta and perform
the t integration in order to obtain that%

\begin{gather}
{\displaystyle\int\limits_{0}^{1}}dt\delta_{\lbrack2n-3]}^{[2n-3]}k\left(
\frac{1}{2}\mathcal{R}ie-t^{2}k^{2}\right)  ^{n-2}=\nonumber\\
\frac{\left(  2n-3\right)  !}{\ell^{2n-3}}\frac{\left(  -1\right)  ^{n-2}%
}{2n-3}+\rho\frac{\left(  -1\right)  ^{n-2}\left(  2n-4\right)  !}{\ell
^{2n-5}}\left[  1-\frac{2n-4}{2n-5}\right]  tr\left[  S^{\left(  0\right)
}\right]  +...\nonumber\\
=\/\left(  -1\right)  ^{n-2}\left[  \frac{\left(  2n-4\right)  !}{\ell^{2n-3}%
}-\rho\frac{\left(  2n-6\right)  !}{2\ell^{2n-5}}\mathcal{R}^{\left(
0\right)  }\right]  +...~.
\end{gather}

Finally, using that to the leading order $\mathcal{R}\left[  \widetilde
{\gamma}\right]  \simeq\rho\mathcal{R}^{\left(  0\right)  }$, where
$\mathcal{R}\left[  \widetilde{\gamma}\right]  $ is the Ricci scalar of the
metric $\widetilde{\gamma}$ and $\mathcal{R}^{\left(  0\right)  }$ is that of
$\sigma^{\left(  0\right)  }$, the expression for $S_{EE}^{ct}$ as given in
Eq.(\ref{Counter-appendix}) can be written entirely in terms of the intrinsic
quantities that depend on $\widetilde{\gamma}$, such that%

\begin{gather}
{\displaystyle\frac{\left(  -1\right)  ^{n+1}}{4G}\frac{\ell^{2n-2}}{\left(
2n-3\right)  !}\int\limits_{0}^{1}}dt\delta_{\lbrack2n-3]}^{[2n-3]}k\left(
\frac{1}{2}\mathcal{R}ie-t^{2}k^{2}\right)  ^{n-2}=\nonumber\\
-\frac{\ell}{4G\left(  2n-3\right)  }+\frac{\ell^{3}R\left[  \widetilde
{\gamma}\right]  }{4G2(2n-3)\left(  2n-4\right)  \left(  2n-5\right)  }+...
\end{gather}
and therefore, we obtain%

\begin{equation}%
\begin{tabular}
[c]{l}%
$S_{EE}^{ct}=-\frac{1}{4G}%
{\displaystyle\int\limits_{\partial\Sigma}}
\frac{d^{2n-3}y\sqrt{\widetilde{\gamma}}\ell}{\left(  2n-3\right)  }+\frac
{1}{4G}%
{\displaystyle\int\limits_{\partial\Sigma}}
\frac{d^{2n-3}y\sqrt{\widetilde{\gamma}}\ell^{3}R\left[  \widetilde{\gamma
}\right]  }{2(2n-3)\left(  2n-4\right)  \left(  2n-5\right)  }+...,$%
\end{tabular}
\end{equation}
which corresponds to Eq.(\ref{S_EE^ren_intrinsic}) of the main text.

\section{Minimal surface for a ball-shaped entangling
region\label{Verification}}

We consider the minimal area condition, as derived in appendix A of
Ref.\cite{AAO1}, which is given by%

\begin{equation}
\partial_{a}\left(  \frac{z_{,a}}{4\rho^{\left(  D-1\right)  /2}m\left(
\rho,x^{a}\right)  }\right)  +\partial_{\rho}\left(  \frac{z_{,\rho}}{\ell
^{2}\rho^{\left(  D-3\right)  /2}m\left(  \rho,x^{a}\right)  }\right)  =0,
\label{Extremal_Area_Condition}%
\end{equation}
where $D$ is the dimension of the AAdS bulk, $z\left(  \rho,x^{a}\right)  $ is
the embedding function of the surface $\Sigma$ expressed in Cartesian
coordinates and%

\begin{equation}
m\left(  \rho,x^{a}\right)  =\sqrt{1+4\frac{\rho}{\ell^{2}}z_{,\rho}z_{,\rho
}+z_{,a}z_{,a}}.
\end{equation}
We also consider the definition of the minimal surface $\Sigma$, which is
given in Eq.(\ref{Minimal_Surface_Param}) of the main text, and corresponds to%

\begin{equation}
\Sigma:\left\{  t=const~;~r^{2}+\ell^{2}\rho=R^{2}\right\}  .
\end{equation}

We proceed to verify that this definition of $\Sigma$ satisfies the minimal
area condition of Eq.(\ref{Extremal_Area_Condition}). Therefore, we first
write the corresponding embedding function $z\left(  \rho,x\right)  $ in
Cartesian coordinates, considering that $r^{2}=z^{2}+x^{a}x^{b}\delta_{ab}$,
for $a,b=1,...,\left(  D-3\right)  $. Thus we have that $z\left(  \rho
,x^{a}\right)  =\pm\sqrt{R^{2}-\ell^{2}\rho-x^{a}x^{b}\delta_{ab}}$. Then,
computing the derivatives of the embedding function, we have that%

\begin{equation}%
\begin{tabular}
[c]{l}%
$z_{,a}=\mp\frac{x^{a}}{\sqrt{R^{2}-\ell^{2}\rho-x^{a}x^{b}\delta_{ab}}%
}~;~z_{,\rho}=\mp\frac{\ell^{2}}{2\sqrt{R^{2}-\ell^{2}\rho-x^{a}x^{b}%
\delta_{ab}}},$\\
$m\left(  \rho,x\right)  =\sqrt{1+4\frac{\rho}{\ell^{2}}z_{,\rho}z_{,\rho
}+z_{,a}z_{,a}}=\frac{R}{\sqrt{R^{2}-\ell^{2}\rho-x^{a}x^{b}\delta_{ab}}},$%
\end{tabular}
\end{equation}
and replacing the corresponding terms into Eq.(\ref{Extremal_Area_Condition}),
we have that%

\begin{equation}%
\begin{tabular}
[c]{l}%
$\partial_{a}\left(  \frac{z_{,a}}{4\rho^{\left(  D-1\right)  /2}m\left(
\rho,x^{a}\right)  }\right)  +\partial_{\rho}\left(  \frac{z_{,\rho}}{\ell
^{2}\rho^{\left(  D-3\right)  /2}m\left(  \rho,x^{a}\right)  }\right)
=\mp\left(  \partial_{a}\left(  \frac{x^{a}}{4\rho^{\left(  D-1\right)  /2}%
R}\right)  +\partial_{\rho}\left(  \frac{1}{2\rho^{\left(  D-3\right)  /2}%
R}\right)  \right)  $\\
$=\mp\left(  \frac{\left(  D-3\right)  }{4\rho^{\left(  D-1\right)  /2}%
R}-\frac{1}{2}\frac{\left(  D-3\right)  }{2}\frac{1}{\rho^{\left(  D-1\right)
/2}R}\right)  =0,$%
\end{tabular}
\end{equation}
which implies that $\Sigma$ is indeed the minimal surface.

\end{document}